\documentclass[aps,prl,twocolumn,showpacs]{revtex4}
\usepackage{bbm,amsmath,amssymb,amsbsy,graphicx,times,subfigure}

\usepackage[latin1]{inputenc}
\renewcommand{\det}{{\rm Det}\,}
\newcommand{\gr}[1]{\boldsymbol{#1}}
\newcommand{\ket}[1]{|#1\rangle}
\newcommand{\bra}[1]{\langle#1|}
\newcommand{\avr}[1]{\langle#1\rangle}
\newcommand{\eq}[1]{Eq.~(\ref{#1})}

\begin{document}
\title{Equivalence between Entanglement and the Optimal Fidelity of
Continuous Variable Teleportation}

\date{August 27, 2005}
\author{Gerardo Adesso}
\author{Fabrizio Illuminati}
\affiliation{Dipartimento di Fisica ``E. R. Caianiello'',
Universit\`a di Salerno, CNR-Coherentia, Gruppo di Salerno, \\ INFN
Sezione di Napoli, Gruppo Collegato di Salerno, Via S. Allende,
84081 Baronissi (SA), Italy}


\begin{abstract}
We devise the optimal form of Gaussian resource states enabling
continuous variable teleportation with maximal fidelity. We
 show that a nonclassical optimal fidelity of $N$-user
teleportation networks is {\it necessary and sufficient} for
$N$-party entangled Gaussian resources, yielding an estimator of
multipartite entanglement. This {\it entanglement of teleportation}
is equivalent to entanglement of formation in the two-user protocol,
and to localizable entanglement in the multi-user one. The
continuous-variable tangle, quantifying entanglement sharing in
three-mode Gaussian states, is operationally linked to the optimal
fidelity of a tripartite teleportation network.
\end{abstract} \maketitle


Quantum teleportation using quadrature entanglement in continuous
variable (CV) systems \cite{review} is in principle
imperfect, due to the impossibility of achieving infinite squeezing.
Nevertheless, by considering the finite quantum correlations between
the quadratures in a two-mode squeezed Gaussian state, a realistic
scheme for CV teleportation was proposed \cite{vaidman,brakim}, and
experimentally implemented to teleport coherent states with a
fidelity  up to ${\cal F} = 0.70 \pm 0.02$ \cite{furuscience}.
Without using entanglement, by purely classical communication, an
average fidelity of ${\cal F}_{cl}=1/2$ is the best that can be
achieved if the alphabet of input states includes all coherent
states with even weight \cite{bfkjmo}. The original teleportation
protocol \cite{brakim} was generalized to a multi-user teleportation
network requiring multipartite CV entanglement in Ref. \cite{bra00}.
This network has been recently  demonstrated experimentally by
exploiting three-mode squeezed Gaussian states, yielding a best
fidelity of ${\cal F} = 0.64 \pm 0.02$ \cite{naturusawa}. The {\em
fidelity}, which quantifies the success of a teleportation
experiment, is defined as ${\cal F} \equiv \bra{\psi^{in}}
\varrho^{out}\ket{\psi^{in}}$, where ``in'' and ``out'' denote the
input and the output state. ${\cal F}$ reaches unity only for a
perfect state transfer, $\varrho^{out} =
\ket{\psi^{in}}\!\bra{\psi^{in}}$. To accomplish teleportation with
high fidelity, the sender (Alice) and the receiver (Bob) must share
an entangled state (resource).
The sufficient fidelity criterion \cite{bfkjmo}
states that, if teleportation is performed with ${\cal F} > {\cal
F}_{cl}$, then the two parties exploited an entangled state. The
converse is generally false, i.e. some entangled resources may yield
lower-than-classical fidelities.

In this Letter we investigate the relation between the fidelity of a
CV teleportation experiment and the entanglement present in the
 resource states. We show that the optimal fidelity,
maximized over all local single-mode operations (at fixed amounts of
noise and entanglement in the resource), is {\it necessary and
sufficient} for the presence of  bipartite (multipartite)
entanglement in  two-mode (multimode) Gaussian resources. Moreover,
it allows for the definition of the {\em entanglement of
teleportation}, an operative estimator of bipartite (multipartite)
entanglement in CV systems. Remarkably, in the multi-user instance,
the optimal shared entanglement is exactly the localizable
entanglement, originally introduced for spin systems \cite{localiz},
which thus acquires for Gaussian states a suggestive operative
meaning in terms of teleportation processes. In the CV scenario,
 a recent study on entanglement sharing led to
the definition of the residual CV tangle, or {\em contangle}
$E_\tau$, as a tripartite entanglement monotone under Gaussian LOCC
for  three-mode Gaussian states \cite{contangle}. This measure too
is here operationally interpreted via the success of a three-party
teleportation network. Besides these fundamental theoretical
results, our findings are of important practical interest, as they
answer the experimental need for the best preparation recipe for
entangled squeezed resources, in order to implement CV teleportation
with the highest  fidelity.

The two-user CV teleportation protocol \cite{brakim} would require,
to achieve unit fidelity, the sharing of an ideal (unnormalizable)
Einstein-Podolski-Rosen (EPR) resource state \cite{epr}, i.e. the
eigenstate of relative position and total momentum of a two-mode
radiation field. An arbitrarily good approximation of the EPR state
is represented by two-mode squeezed Gaussian states with squeezing
parameter $r \rightarrow \infty$. In a CV system consisting of $N$
canonical bosonic modes, and described by the vector $\hat{X} =
\{\hat x_1, \hat p_1, \ldots, \hat x_N, \hat p_N\}$ of the field
quadrature operators \cite{noteunit}, Gaussian states (such as
thermal, coherent, squeezed states) are fully characterized by the
first statistical moments (arbitrarily adjustable by local
unitaries: we will set them to zero) and by the $2N \times 2N$
covariance matrix (CM) $\gr{\sigma}$ of the second moments
$\sigma_{ij}=1/2\langle\{\hat{X}_i,\hat{X}_j\}\rangle$. A two-mode
squeezed state can be, in principle, produced by mixing a
momentum-squeezed state and a position-squeezed state, with
squeezing parameters $r_1$ and $r_2$ respectively, through a 50:50
ideal (lossless) beam splitter. In practice, due to experimental
imperfections and unavoidable thermal noise the two initial squeezed
states will be mixed. To perform a realistic analysis, we must then
consider two thermal squeezed single-mode states \cite{notemix},
described by the following quadrature operators in Heisenberg
picture \vspace*{-.1cm}
\begin{eqnarray}
  \hat x_1^{sq} \!&=&\! \sqrt{n_1} e^{r_1} \hat x_1^0\,,\quad
  \hat p_1^{sq} = \sqrt{n_1} e^{-r_1} \hat p_1^0\,, \label{momsq}  \\
   \hat x_2^{sq} \!&=&\! \sqrt{n_2} e^{-r_2} \hat x_2^0\,,\quad
  \hat p_2^{sq} = \sqrt{n_2} e^{r_2} \hat p_2^0\,, \label{possq}
\vspace*{-.2cm}\end{eqnarray} where the suffix ``0'' refers to the
vacuum. The action of an ideal (phase-free) beam splitter operation
on a pair of modes $i$ and $j$ is defined as $
\hat{B}_{i,j}(\theta):\left\{
\begin{array}{l}
\hat a_i \rightarrow \hat a_i \cos\theta + \hat a_j\sin\theta \\
\hat a_j \rightarrow \hat a_i \sin\theta - \hat a_j\cos\theta \\
\end{array} \right.$,
where $\hat a_k = (\hat x_k + i \hat p_k)/2$ is the annihilation
operator of mode $k$.  When applied to the two modes of
Eqs.~(\ref{momsq},\ref{possq}), the beam splitter entangling
operation ($\theta = \pi/4$) produces a symmetric mixed state
\cite{refbs}, depending on the squeezings $r_{1,2}$ and on the
thermal noises $n_{1,2}$. The noise can be difficult to control and
reduce in the lab, but it is quantifiable. Now, keeping $n_1$ and
$n_2$ fixed, all states produced starting with different $r_1$ and
$r_2$, but with equal average $\bar r \equiv (r_1 + r_2)/2$, are
completely equivalent up to local unitary operations and possess, by
definition, the same entanglement. Let us recall that a two-mode
Gaussian state is entangled if and only if it violates the
positivity of partial transpose (PPT) condition $\eta \ge 1$
\cite{simon}. The quantity $\eta$ is the smallest symplectic
eigenvalue of the partially transposed CM, which is obtained from
the CM of the Gaussian state by performing trasposition (time
reversal in phase space \cite{simon}) in the subspace associated to
either one of the modes. The CM $\gr\sigma$ of a generic two-mode
Gaussian state can be written in the block form $ \gr\sigma =
{\footnotesize \left(
\begin{array}{cc}
\gr \alpha & \gr \gamma \\
\gr\gamma^{\sf T} & \gr \beta \\
\end{array}%
\right)}, $ where $\gr \alpha$ and $\gr \beta$ are the CM's of the
individual modes, while the matrix $\gr \gamma$ describes intermodal
correlations. One then has  $2 \eta^2 = \Sigma(\gr \sigma) -
\sqrt{\Sigma^2(\gr \sigma)-4\det{\gr \sigma}}$, where $\Sigma(\gr
\sigma) \equiv \det\gr\alpha + \det\gr\beta - 2 \det\gr\gamma$
\cite{extremal}. The parameter $\eta$ also provides a quantitative
characterization of CV entanglement, because the logarithmic
negativity and, equivalently for symmetric states
($\det\gr\alpha=\det\gr\beta$), the entanglement of formation $E_F$,
are both decreasing functions of $\eta$. For symmetric Gaussian
states the bipartite entanglement $E_F$ reads \cite{efprl}
\vspace*{-.1cm}
\begin{equation}\label{eof}
E_F (\gr \sigma) = \max \{0,\, f(\eta)\},
\vspace*{-.1cm}\end{equation}
 with $\quad f(x) \equiv \frac{(1+x)^2}{4x}
\log{\frac{(1+x)^2}{4x}}
- \frac{(1-x)^2}{4x} \log{\frac{(1-x)^2}{4x}}$.\\

\vspace*{-.2cm}

For the mixed two-mode states considered here, we have
\vspace*{-.1cm}\begin{equation}\label{eta}
 \eta = \sqrt{n_1 n_2
}e^{-(r_1+r_2)}\,.\vspace*{-.1cm}
\end{equation}
The entanglement thus depends both on the arithmetic mean of the
individual squeezings, and on the geometric mean of the individual
noises, which is related to the purity of the state $\mu = (n_1
n_2)^{-1}$. The teleportation success, instead, depends separately
on each of the four single-mode parameters. The fidelity (averaged
over the complex plane) for teleporting an unknown single-mode
coherent state can be computed by writing the quadrature operators
in  Heisenberg picture \cite{bra00,vanlok}:
\begin{equation}\label{fid}
{\cal F} \equiv \phi ^{-1/2},\:\: \phi = \left\{\left[\avr{(\hat
x_{tel})^2}+1\right]\left[\avr{(\hat
p_{tel})^2}+1\right]\right\}/4\,,
\end{equation}
where $\avr{(\hat x_{tel})^2}$ and $\avr{(\hat p_{tel})^2}$ are the
variances of the canonical operators $\hat x_{tel}$ and $\hat
p_{tel}$ which describe the teleported mode. For the utilized
states, we have $ \hat x_{tel} = \hat x^{in} - \sqrt{2 n_2} e^{-r_2}
\hat x_2^0\,,\:\:\hat p_{tel} = \hat p^{in} + \sqrt{2 n_1} e^{-r_1}
\hat p_1^0\,,$ where the suffix ``in'' refers to the input coherent
state to be teleported. Recalling that, in our units
\cite{noteunit}, $\avr{(\hat x_{i}^0)^2}=\avr{(\hat
p_{i}^0)^2}=\avr{(\hat x^{in})^2}=\avr{(\hat p^{in})^2}=1$, we can
 compute the fidelity from \eq{fid}, obtaining
$\phi(r_{1,2},n_{1,2}) =
e^{-2(r_1+r_2)}(e^{2r_1}+n_1)(e^{2r_2}+n_2)\,.$ It is convenient to
replace $r_1$ and $r_2$ by $\bar r$ and $d \equiv
(r_1-r_2)/2$:\vspace*{-.1cm}
\begin{equation}\label{fidd}
\phi(\bar r,d,n_{1,2}) = e^{-4 \bar r}(e^{2(\bar r +
d)}+n_1)(e^{2(\bar r - d)}+n_2)\,. \vspace*{-.1cm}\end{equation}
Maximizing the fidelity for given entanglement and noises of the
Gaussian resource state (i.e.~for fixed $n_{1,2},\bar r$) simply
means finding the $d=d^{opt}$ which minimizes the quantity $\phi$ of
\eq{fidd}. Being $\phi$ a convex function of $d$, it suffices to
find the zero of $\partial \phi/\partial d$, yielding $d^{opt} =
\frac14 \log{\frac{n_1}{n_2}}$. For equal noises, $d^{opt}=0$,
indicating that the best preparation of the entangled resource state
needs two equally squeezed single-mode states, in agreement with the
results presented in Ref. \cite{bowen} for pure states. For
different noises, however, the optimal procedure involves two
different squeezings such that $r_1 - r_2 = 2 d^{opt}$. Inserting
$d^{opt}$ in \eq{fidd} we have the optimal fidelity \vspace*{-.2cm}
\begin{equation}\label{fiopt2}
{\cal F}^{opt} = 1/({1+\eta})\,, \vspace*{-.2cm}\end{equation} where
$\eta$ is exactly the lowest symplectic eigenvalue of the partial
transpose, defined by \eq{eta}. \eq{fiopt2} clearly shows that the
optimal teleportation fidelity depends only on the entanglement of
the resource state, and vice versa. In fact, the fidelity criterion
becomes {\em necessary and sufficient} for the presence of the
entanglement, if ${\cal F}^{opt}$ is considered: the optimal
fidelity is classical  for $\eta \ge 1$ (separable state) and
greater than the classical threshold for any entangled state.
Moreover, ${\cal F}^{opt}$ provides a quantitative measure of
entanglement completely equivalent to the two-mode entanglement of
formation, namely (from Eqs.~(\ref{eof},\ref{fiopt2})): $E_F = \max
\{0,\, f(1/{\cal F}^{opt}-1)\}$. In the limit of infinite squeezing
($\bar r \rightarrow \infty$), ${\cal F}^{opt}$ reaches $1$ for any
amount of finite thermal noise. On the other extreme, due to the
convexity of $\phi$, the lowest fidelity (maximal waste of
entanglement) is attained at one of the boundaries $d = \pm \bar r$,
meaning that one of the squeezings $r_{1,2}$ vanishes.
For infinite squeezing, the worst fidelity cannot exceed
$1/\sqrt{\max\{n_1,n_2\}}$, falling below $1/2$ for strong enough
noise.

We now extend our analysis to a quantum teleportation-network
protocol, involving $N$ users who share a genuine $N$-partite
entangled Gaussian resource, completely symmetric under permutations
of the modes \cite{bra00}. Two parties are randomly chosen as sender
(Alice) and receiver (Bob), but this time, in order to accomplish
teleportation of an unknown coherent state, Bob needs the results of
$N-2$ momentum detections performed by the other cooperating
parties. A nonclassical teleportation fidelity (i.e. ${\cal F} >
{\cal F}^{cl}=1/2$) between {\em any} pair of parties is sufficient
for the presence of genuine $N$-partite entanglement in the shared
resource, while in general the converse is false (see {\it
e.g.}~Fig.1 of Ref.~\cite{bra00}). Our aim is to determine the
optimal multi-user teleportation fidelity, and to extract from it a
quantitative information on the multipartite entanglement in the
shared resources. We begin by considering a mixed momentum-squeezed
state described by $r_1,n_1$ as in \eq{momsq}, and $N-1$
position-squeezed states of the form \eq{possq}. We then combine the
$N$ beams into an $N$-splitter \cite{bra00}: $ \hat{N}_{1\ldots N}
\equiv \hat{B}_{N-1,N}(\pi/4)\hat{B}_{N-2,N-1}
(\cos^{-1}1/\sqrt{3})\cdot
\ldots\cdot\hat{B}_{1,2}(\cos^{-1}1/\sqrt{N}).$ The resulting state
is a completely symmetric mixed Gaussian state of a $N$-mode CV
system, parametrized by $n_{1,2}$, $\bar r$ and $d$. Once again, all
states with equal $\{n_{1,2},\bar r\}$ belong to the same
iso-entangled class of equivalence. For $\bar r \rightarrow \infty$
and for $n_{1,2}=1$ (pure states), these states reproduce the
(unnormalizable) CV Greenberger-Horne-Zeilinger (GHZ) \cite{ghz}
state $\int dx \ket{x,x,\ldots,x}$, an eigenstate with total
momentum zero and all relative positions $x_i - x_j = 0$
($i,j=1,\ldots,N$).  Choosing randomly two modes, denoted by the
indices $k$ and $l$, to be respectively the sender and the receiver,
the teleported mode is described by the following quadrature
operators (see Refs. \cite{bra00,vanlok} for further details): $\hat
x_{tel} = \hat x_{in} - \hat x_{rel},\:\: \hat p_{tel} = \hat p_{in}
+ \hat p_{tot},$ with $ \hat x_{rel} = \hat x_{k} - \hat x_{l}$ and
$\hat p_{tot} = \hat p_{k} + \hat p_{l} + g_N \sum_{j \ne k,l} {\hat
p_{j}}\,, $ where $g_N$ is an experimentally adjustable gain. To
compute the teleportation fidelity from \eq{fid}, we need the
variances of $\hat x_{rel}$ and $\hat p_{tot}$. From the action of
the $N$-splitter, we have \vspace*{-.1cm}
\begin{eqnarray}\label{avrel}
  \avr{(\hat x_{rel})^2} &=& 2 n_2 e^{-2 (\bar r - d)}\,, \nonumber\\
  \avr{(\hat p_{tot})^2} &=& \big\{ [2+ (N-2) g_N]^2 n_1 e^{-2(\bar r + d)}\\
  &+& 2 [g_N-1]^2(N-2) n_2 e^{2(\bar r - d)}  \big\}/4\,. \nonumber
\vspace*{-.2cm}\end{eqnarray}

\begin{figure}[t!]
\includegraphics[height=4cm]{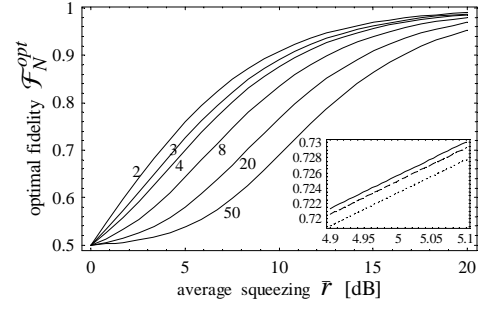}
\caption{Optimal teleportation fidelity of coherent states from any
sender to any receiver chosen from $N$ (= 2, 3, 4, 8, 20, and 50)
parties, using pure $N$-party entangled symmetric Gaussian
resources. The optimal fidelity is nonclassical for any $N$, if the
initial squeezings are adjusted as in  \eq{dopt}. At fixed
entanglement, states produced with all equal squeezers yield
nonclassical fidelities for $N \ge 30$ (see Fig.1 of Ref.
\cite{bra00}). In the inset we compare, for $N=3$ and a window of
average squeezing, the optimal fidelity (solid line), the fidelity
with unbiased states \cite{bowen} (dashed line), and the fidelity
with equally squeezed states \cite{bra00} (dotted line). The three
curves are  close, but the optimal preparation yields always the
highest fidelity.} \label{finopt}
\end{figure}

The optimal fidelity can  be found in two straightforward steps: 1)
minimizing $\avr{(\hat p_{tot})^2}$ with respect to $g_N$ (i.e.
finding the optimal gain $g_N^{opt}$); 2) minimizing the resulting
$\phi$ with respect to $d$ (i.e. finding the optimal $d_N^{opt}$).
The results are\vspace*{-.1cm}
\begin{eqnarray}
  g_N^{opt} &=& 1- N/\left[(N-2)+2e^{4 \bar r} n_2/n_1\right]\,, \label{gopt}\\
  d_N^{opt} &=& \bar r + \log\left\{ N/\left[(N-2)+2e^{4 \bar r} n_2/n_1\right]
  \right\}/4\,. \vspace*{-.2cm}\label{dopt}
\end{eqnarray}
Inserting Eqs.~(\ref{avrel}--\ref{dopt}) in \eq{fid}, we find the
optimal teleportation-network fidelity, which can be put in the
following general form for $N$ modes \vspace*{-.2cm}
\begin{equation}\label{fidn}
{\cal F}_N^{opt} = \frac{1}{1+\eta_N}\,,\quad \eta_N \equiv
\sqrt{\frac{N n_1 n_2}{2e^{4 \bar r} +(N-2)n_1/n_2}}\,.
\vspace*{-.1cm}\end{equation} For $N=2$, $\eta_2 = \eta$ from
\eq{eta}, showing that the general multipartite protocol comprises
the standard bipartite case. By comparison with \eq{fiopt2}, we
observe that, for any $N>2$, the quantity $\eta_N$ plays the role of
a generalized  symplectic eigenvalue, whose physical meaning will be
clear soon. Before that, it is worth commenting on the form of the
optimal resources, focusing for simplicity on the pure-state setting
($n_{1,2}=1$). The optimal form of the shared $N$-mode symmetric
Gaussian states, for $N>2$, is neither unbiased in the $x_i$ and
$p_i$ quadratures (like the states discussed in Ref. \cite{bowen}
for three modes), nor constructed by $N$ equal squeezers ($r_1=r_2=
\bar r$). This latter case, which has been implemented
experimentally for $N=3$ \cite{naturusawa}, is clearly not optimal,
yielding fidelities lower than $1/2$ for $N\ge30$ and $\bar r$
falling in a certain interval \cite{bra00}. The explanation of this
paradoxical behaviour, provided by the authors of Ref. \cite{bra00},
is that their teleportation scheme might not be optimal. Our
analysis shows instead that the problem does not lie in the
protocol, but rather in the employed states. If the shared $N$-mode
resources are prepared (or locally transformed) in the optimal form
of \eq{dopt}, the teleportation fidelity is guaranteed to be
nonclassical (see Fig.\ref{finopt}) as soon as $\bar r>0$ for any
$N$, in which case the considered class of pure states is genuinely
multiparty entangled \cite{vanlok,matrioska}. Therefore {\em a
nonclassical optimal fidelity is  necessary and sufficient for the
presence of multipartite entanglement in any multimode symmetric
Gaussian state}, shared as a resource for CV teleportation. On the
opposite side, the worst preparation scheme of the multimode
resource states, even retaining the optimal protocol
($g_N=g_N^{opt}$), is obtained setting $r_1=0$ if $n_1 > 2 n_2 e^{2
\bar r}/(N e^{2 \bar r} + 2 - N)$, and $r_2=0$ otherwise. For equal
noises ($n_1=n_2$), the case $r_1=0$ is always the worst one, with
asymptotic fidelities (in the limit $\bar r \rightarrow \infty$)
equal to $1/\sqrt{1+N n_{1,2}/2}$, so rapidly dropping with $N$ at
given noise.

The meaning of $\eta_N$, crucial for the quantification of the
multipartite entanglement, stems from the following argument. The
teleportation network \cite{bra00} is realized in two steps: first,
the $N-2$ cooperating parties perform local measurements on their
modes, then Alice and Bob exploit their resulting highly entangled
two-mode state to accomplish teleportation. Stopping at the first
stage, the protocol describes a concentration, or {\em localization}
of the original $N$-partite entanglement, into a bipartite two-mode
entanglement \cite{bra00,vanlok}. The maximum entanglement that can
be concentrated on a pair of parties by locally measuring the
others, is known as the {\em localizable entanglement} (LE) of a
multiparty system \cite{localiz}. Here, the LE is the maximal
entanglement concentrable onto two modes, by unitary operations and
nonunitary momentum detections performed locally on the other $N-2$
modes. The two-mode entanglement of the resulting state (described
by a CM $\gr \sigma_{loc}$) is quantified in terms of the symplectic
eigenvalue $\eta_{loc}$ of its partial transpose. Due to the
symmetry of the original state and of the protocol (the gain is the
same for every mode), the localized two-mode state is symmetric too.
It has been proven \cite{extremal} that, for two-mode symmetric
Gaussian states, the symplectic eigenvalue $\eta$ is related to the
EPR correlations by the expression $4 \eta =\avr{(\hat x_1 - \hat
x_2)^2} + \avr{(\hat p_1 + \hat p_2)^2}$. For the state
$\gr\sigma_{loc}$, this means $4\eta_{loc} = \avr{(\hat x_{rel})^2}
+ \avr{(\hat p_{tot})^2}$, where the variances have been computed in
\eq{avrel}. Minimizing $\eta_{loc}$  with respect to $d$ means
finding the optimal set of local unitary operations (unaffecting
multipartite entanglement) to be applied to the original multimode
mixed resource described by $\{n_{1,2},\bar r, d\}$; minimizing then
$\eta_{loc}$ with respect to $g_N$ means finding the optimal set of
momentum detections to be performed on the transformed state in
order to localize the highest entanglement on a pair of modes. From
\eq{avrel}, the optimizations are readily solved and yield the same
optimal $g_N^{opt}$ and $d_N^{opt}$ of Eqs.~(\ref{gopt},\ref{dopt}).
The resulting two-mode state contains a localized entanglement {\em
exactly} quantified by the quantity $\eta_{loc}^{opt} = \eta_N$. It
is now clear that $\eta_N$ of \eq{fidn} is a proper symplectic
eigenvalue, being the smallest one of the partial transpose of the
optimal two-mode state that can be extracted from a $N$-party
entangled resource by local measurements on the remaining modes.
\eq{fidn} thus provides a bright connection between two {\em
operative} aspects of  multipartite entanglement in CV systems:  the
maximal fidelity achievable in a multi-user teleportation network
\cite{bra00}, and the LE \cite{localiz}.

This results yield quite naturally a direct operative way to
quantify multipartite entanglement in $N$-mode (mixed) symmetric
Gaussian states, in terms of the so-called {\em Entanglement of
Teleportation}, defined as the normalized optimal fidelity
\vspace*{-.2cm}
\begin{equation}\label{et}
E_T \equiv \max\left\{0,\frac{{\cal F}_N^{opt}-{\cal
F}_{cl}}{1-{\cal
F}_{cl}}\right\}=\max\left\{0,\frac{1-\eta_N}{1+\eta_N}\right\}\,,
\vspace*{-.2cm}\end{equation} and thus ranging from 0 (separable
states) to 1 (CV GHZ state). A homonym but different concept has
also been introduced for discrete variables \cite{rigolazzo}. The
localizable entanglement of formation $E_F^{loc}$ of $N$-mode
symmetric Gaussian states is a monotonically increasing function of
$E_T$, namely: $E_F^{loc} = f[(1-E_T)/(1+E_T)]$, with $f(x)$ defined
after \eq{eof}. For $N=2$ the state is already localized and
$E_F^{loc} = E_F$.

Remarkably for three-mode pure (symmetric) Gaussian states, the
residual contangle $E_\tau$, a tripartite entanglement monotone
under Gaussian LOCC that quantifies CV entanglement sharing
\cite{contangle}, is also a monotonically increasing function of
$E_T$, thus providing another {\em equivalent} quantitative
characterization of
genuine tripartite CV entanglement. In formula: \\
{\hspace*{.2cm} $E_\tau = \log^2{\frac{2\sqrt2
E_T-(E_T+1)\sqrt{E_T^2+1}}{(E_T-1)\sqrt{E_T(E_T+4)+1}}}-\frac12
\log^2{\frac{E_T^2+1}{E_T(E_T+4)+1}}\,.$}\\
This finding suggests an experimental test, in terms of optimal
fidelities in teleportation networks \cite{naturusawa}, to verify
the promiscuous sharing of tripartite CV entanglement in pure
symmetric
three-mode Gaussian states, discovered in Ref. \cite{contangle}.\\
\indent
 Whether an expression of the form \eq{et} connecting $E_T$ to the
symplectic eigenvalue $\eta_N$ remains true for generalized
teleportation protocols \cite{fiura} and for nonsymmetric entangled
resources, is currently an open question. However, nonsymmetric
Gaussian states are never optimal candidates for communication
protocols, as their maximum achievable entanglement decreases with
increasing asymmetry \cite{extremal}, and therefore they are
automatically ruled out by the present analysis.

This work is supported by MIUR, INFN, CNR. GA thanks M. Santos, A.
Serafini and P. van Loock for discussions.


\begin{thebibliography}{99}


\bibitem{review} S.\,L.\,Braunstein\,and\,P.\,van\,Loock,\,Rev.\,Mod.\,Phys.\,{\bf 77},\,513\,(2005).
\bibitem{vaidman} L. Vaidman, Phys. Rev. A {\bf 49}, 1473 (1994).

\bibitem{brakim} S.\,L.\,Braunstein\,and\,H.\,J.\,Kimble,\,Phys.\,Rev.\,Lett.\,{\bf 80},\,869\,(1998).

\bibitem{furuscience} A. Furusawa {\em et al.}, Science {\bf 282},
706 (1998); W. P. Bowen~{\em et~al.}, Phys. Rev. A {\bf 67}, 032302
(2003); N. Takei {\em et al.}, Phys. Rev. Lett. \textbf{94}, 220502
(2005).


\bibitem{bfkjmo} S. L. Braunstein {\em et al.}, J.
Mod. Opt. {\bf 47}, 267 (2000); \\ K. Hammerer {\em et al.}, Phys.
Rev. Lett. {\bf 94}, 150503 (2005).


\bibitem{bra00} P.\,van\,Loock\,and\,S.\,L.\,Braunstein,\,Phys.\,Rev.\,Lett.\,{\bf 84},\,3482\,(2000).


\bibitem{naturusawa} H.\,Yonezawa,\,T.\,Aoki,\,and\,A.\,Furusawa,\,Nature\,{\bf 431},\,430\,(2004).

\bibitem{localiz} F. Verstraete, M. Popp, and J. I. Cirac, Phys.
Rev. Lett. {\bf 92}, 027901 (2004); M. Popp {\em et al.}, Phys. Rev.
A {\bf 71}, 042306 (2005).

\bibitem{contangle} G. Adesso and F. Illuminati, quant-ph/0410050v3.



\bibitem{epr} A.\,Einstein,\,B.\,Podolsky,\,and\,N.\,Rosen,\,Phys.\,Rev.\,{\bf 47},\,777\,(1935).

\bibitem{noteunit} Defining the annihilation operator $\hat a_i =
(\hat x_i + i \hat p_i)/2$ for mode $i$, the Weyl algebra $[\hat
a_i,\,\hat a_i^\dagger]=1$ implies $[\hat x_i,\,\hat p_i]=2i$, so
$\hbar=2$.


\bibitem{notemix} Any losses due to  optical elements and/or to
propagating beams can be embedded into the initial single-mode noise
factors.

\bibitem{refbs} M. S. Kim {\em et al.}, Phys. Rev. A {\bf 65},
032323 (2002); \\A. Serafini {\em et al.}, {\em ibid.} {\bf 69},
022318 (2004).


\bibitem{simon} R. Simon, Phys. Rev. Lett. {\bf 84}, 2726 (2000).


\bibitem{extremal} G. Adesso, A. Serafini, and F. Illuminati, Phys. Rev. A {\bf 70}, 022318 (2004).


\bibitem{efprl} G. Giedke {\it et al.}, Phys. Rev. Lett. {\bf 91}, 107901
(2003).

\bibitem{vanlok} P. van Loock, Fortschr. Phys. {\bf 50}, 12 1177 (2002).

\bibitem{bowen}  W. P. Bowen {\em et al.}, J. Mod.
Opt. {\bf 50}, 801 (2003).


\bibitem{ghz} D. M. Greenberger {\em et al.}, Am. J. Phys. {\bf 58}, 1131 (1990).

\bibitem{matrioska} G. Adesso, A. Serafini, and F. Illuminati, Phys.
Rev. Lett. {\bf 93}, 220504 (2004).


\bibitem{rigolazzo} G. Rigolin, Phys. Rev. A {\bf 71}, 032303 (2005).

\bibitem{fiura} J. Fiur\'a\u sek, Phys. Rev. A {\bf 66}, 012304
(2002); S. Pirandola, S. Mancini, and D. Vitali, {\em ibid.} {\bf
71}, 042326 (2005).


\end{thebibliography}
\end{document}